\documentclass[12pt,preprint]{aastex}

\def\ltsima{$\; \buildrel < \over \sim \;$}
\def\gtsima{$\; \buildrel > \over \sim \;$}
\def\lsim{\lower.5ex\hbox{\ltsima}}
\def\gsim{\lower.5ex\hbox{\gtsima}}
\def\lapp{\ifmmode\stackrel{<}{_{\sim}}\else$\stackrel{<}{_{\sim}}$\fi}
\def\gapp{\ifmmode\stackrel{>}{_{\sim}}\else$\stackrel{<}{_{\sim}}$\fi}

\newdimen\minuswidth    
\setbox0=\hbox{$-$}
\minuswidth=\wd0
\catcode`@=\active
\def@{\kern\minuswidth}
\newdimen\digitwidth    
\setbox0=\hbox{\rm0}
\digitwidth=\wd0
\catcode`!=\active
\def!{\kern\digitwidth}
 
\begin{document} 
 
\title{The puzzling dynamical status of the core of the globular
cluster NGC~6752
 \footnote{Based on observations with the NASA/ESA HST,
obtained at the Space Telescope Science Institute, which is operated
by AURA, Inc., under NASA contract NAS5-26555. Also based on WFI
observations collected at the European Southern Observatory,
La Silla, Chile, within the observing
programmes 62.L-0354 and 64.L-0439.}}
 
\author{F.~R.~Ferraro\altaffilmark{1},
A.~Possenti\altaffilmark{2,3}, 
E.~Sabbi\altaffilmark{1},
P.~Lagani\altaffilmark{1},
R.~T.~Rood\altaffilmark{4}, 
N.~D'Amico\altaffilmark{3,5},
L.~Origlia\altaffilmark{2}}
\medskip
\affil{\altaffilmark{1}Dipartimento di Astronomia, Universit\`a di
Bologna, via Ranzani 1, I--40126 Bologna, Italy, ferraro@bo.astro.it}
\affil{\altaffilmark{2}INAF-Osservatorio Astronomico di Bologna, via
Ranzani 1, I--40126 Bologna, Italy}
\affil{\altaffilmark{3}INAF-Osservatorio Astronomico di Cagliari,
Loc. Poggio dei Pini, Strada 54, I--09012 Capoterra, Italy}
\affil{\altaffilmark{4}Astronomy Dept., University of Virginia,
Charlottesville, USA}
\affil{\altaffilmark{5}Dipartimento di Fisica, Universit\`a di Cagliari,
Strada Provinciale Monserratu--Sestu, km 0.700, I--09042 Monserrato, Italy}
 
\begin{abstract}
We have used high resolution {\it WFPC2-HST} and ground based Wide
Field images to determine the center of gravity and construct an
extended radial density and brightness profile of the cluster NGC~6752
including, for the first time, detailed star counts in the very inner
region. The barycenter of the 9 innermost X-ray sources detected by
{\it Chandra} is located only $1\farcs 9$ off the new center of
gravity.  Both the density and the brightness profile of the central
region are best fitted by a double King model, suggesting that NGC~6752
is experiencing a post-core collapse ``bounce''.  Taking advantage
from our new optical data, we discuss the puzzling nature of the
accelerations displayed by the innermost millisecond pulsars detected
in this cluster. We discuss two possible origins to the accelerations:
1) the overall cluster gravitational potential which would require a
central projected mass to light ratio of order 6--7 and the existence
of a few thousand solar masses of low-luminosity matter within the
inner $0.08$ pc of NGC~6752; 2) the existence of a local perturber(s)
of the pulsar dynamics, such as a recently proposed binary black hole
of intermediate (100--$200\,M_\odot$) mass.
\end{abstract}


\section{Introduction} 
\label{intro}
Although NGC~6752 is one of the nearest clusters, there is no
consensus in the literature on its dynamical status. It was initially
classified as a post-core collapse (PCC) by Djorgovski \& King (1986,
hereafter DK86) and Auriere \& Ortolani (1989, hereafter AO89). 
Later Lugger, Cohn \& Grindlay (1995, hereafter LCG95) argued that the
radial profile was not inconsistent with a King model.
As part of a project devoted to the study the characteristics of globulars
harboring millisecond pulsars (MSPs), we present a new determination
of the center of gravity (\S 3) and of the star density profile (\S 4)
of NGC~6752 both based on new high-resolution and wide-field
observations. By coupling the superior resolution of HST with
the imaging capability of the wide field camera (WFI) mounted at the
2.2m ESO-MPI telescope, we have obtained the most accurate and
extended radial profile ever published for this cluster.  After
modeling these data (\S 5) we examine the dynamical status of the
cluster (\S 6).

Recently, observations of $|\dot{P}/P|$ and location of five
millisecond pulsars in NGC~6752 (D'Amico et al. 2002) have suggested a
surprisingly high mass-to-light ratio ${\cal{M}}/{\cal{L}}_V$ in its
core and the occurrence of non thermal dynamics in the inner regions
(see Colpi, Possenti \& Gualandris 2002). The determination of
${\cal{M}}/{\cal{L}}_V$ by D'Amico et al. relied on results of pulsar
timing and published optical data derived from medium resolution
ground based observation only.  Taking advantage from our new optical
data, we re-examine in \S 7 the D'Amico et al. measurement, discussing
the possibile origin and the consequences of the observed values of
$|\dot{P}/P|$ at some length.

\section{Observations and  data analysis}
\label{obs}
The photometric data used here consist of two sets: {\it (i)---High
resolution set}---a series of HST- WFPC2 images was
obtained in March 2001, through the F555W ($V$) and F336W ($U$)
filters as part of a long term project (GO-8709, PI: F. R. Ferraro)
aimed to study the central stellar populations in a set of Galactic
Globular Glusters (GGCs).  In this dataset the planetary camera (PC,
which has the highest resolution $\sim 0\farcs{046}/{\rm pixel}$) was
roughly centered on the cluster center while the Wide Field (WF)
cameras (at lower resolution $\sim 0\farcs{1}/{\rm pixel}$) sampled
the surrounding outer regions; {\it (ii)---Wide Field set---}a
complementary set of multi-filter ($B$, $V$, $I$) wide field images
was secured during an observing run at the 2.2m ESO-MPI telescope at
ESO (La Silla) in July 1999, using the Wide Field Imager (WFI).  The
WFI has exceptional imaging capabilities---the image consists of a
mosaic of 8 CCD chips (each with a field of view of $8'\times 16'$)
giving a global field of view of $33'\times 34'$.  The cluster was
roughly centered on chip $\#2$.

\subsection{Photometry}  
\label{photometry}
The raw WFI images were corrected for bias and flat field, and the
over-scan region was trimmed using standard IRAF\footnote{IRAF is
distributed by the National Optical Astronomy Observatory, which is
operated by the Association of Universities for Research in Astronomy,
Inc., under cooperative agreement with the National Science
Fundation.} tools.  The point spread function (PSF) fitting procedure
was performed independently on each $V$ and $B$ images, using DAOPHOT
II (Stetson 1994).  A final catalog listing the instrumental $B,~V$
magnitudes for all the stars in each field has been obtained by
cross-correlating the $B$ and $V$ catalogs.  The WFI catalog was
finally calibrated by using the data-set of Buonanno et al. (1986).
 
The photometric reductions of the high resolution images were carried
out using ROMAFOT (Buonanno et al. 1983\nocite{b+83}), a package
developed to perform accurate photometry in crowded fields and
specifically optimized to handle under-sampled PSF (Buonanno \&
Iannicola 1989\nocite{bi89}) as in the case of the HST-WF chips. 

PSF-fitting instrumental magnitudes have been obtained using the
standard procedure described in Ferraro et al. (1997a, 2001).  The
final catalog of the F555W and F336W magnitudes was calibrated by
using the zero-points listed by Holtzman et al. (1995).

\subsection{Astrometry}
\label{astrometry}
The recently released {\it Guide Star Catalog} ($GSCII$) 
was used to search for astrometric standards in the entire
WFI image field of view. Several hundred astrometric $GSCII$ reference
stars were found in each chip, allowing an accurate absolute
positioning of the sources.  An astrometric solution has been obtained
for each of the 8 WFI chips independently, by using a procedure
developed at the Bologna Observatory (Montegriffo et al.  2002, in
preparation).  At the end of the entire procedure, the rms residuals
were of the order of $\sim 0\farcs2$ both in RA and Dec.

The small field ($2\farcm5$ on the side) of the high resolution WFPC2/HST
images was entirely contained within the field of view of the WFI chip
$\#2$, which imaged the central part of the cluster.  Thus we used
more than 1200 bright stars in the WFI catalog lying in the WFPC2-FoV
as {\it secondary astrometric standards} in order to properly find an
astrometric solution for the WFPC2 catalog. We estimate that the
global uncertainty in the astrometric procedure is less than $\sim
0\farcs4$ both in RA and Dec.  At the end of the procedure the two
catalogs (WFPC2 and WFI) have a fully homogeneous absolute coordinate
system.

\begin{figure}
\plotone{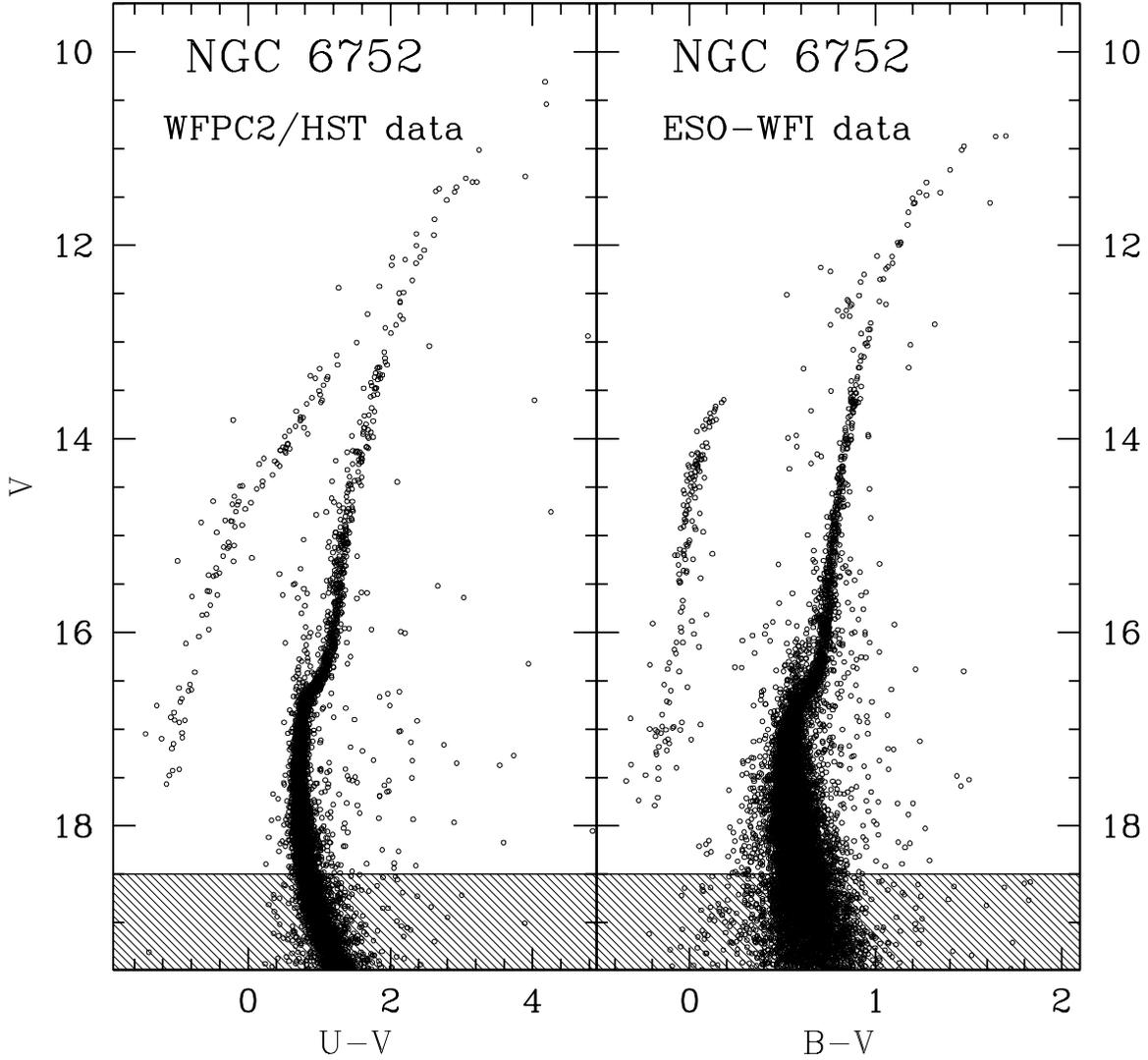} 
\caption{\label{CMD} \footnotesize{CMDs
for stars in the two catalogs.  {\it Panel (a):} The high resolution
HST catalog in the $({\rm F555W,~F336W-F555W})$ plane; {\it Panel
(b):} The wide-field WFI catalog in the $(V,~B-V)$ plane.  Only stars
with $r>120\arcsec$ from the cluster center are plotted. Stars with $V>18.5$
(dashed region) have been not considered in the density profile
construction.}}
\end{figure}

The two lists were then matched together: stars in the overlapping
area were used to recheck the homogeneity of the calibration of
the $V$-magnitude (which is used to construct the surface profile) in
the two catalogs. To do this the HST F555W band was transformed into
the Johnson $V$ system and a homogeneous list of $V$ magnitudes and
absolute coordinates for the sources in the HST and the WFI catalogs
was produced.  In order to avoid strong selection effects due to
possible incompleteness of the samples (in particular the WFI catalog),
we grouped the two samples as follows: {\sl (1) HST sample:} all stars in
the WFPC2 FoV with $V<18.5$ and $r<96\arcsec$ from the cluster center;
{\sl (2) WFI sample:} all the stars in the WFI FoV with $V<18.5$ and
$r>120\arcsec$ from the cluster center.  The Color-Magnitude Diagrams
(CMDs) derived from these samples are shown in Figure~\ref{CMD}.
  
\section{The center of gravity}  
\label{center} 
The first step toward the computation of the density profile 
is the determination of the
center of gravity $C_{\rm grav}$ of the cluster. In doing this, we
estimated the position of the geometrical center of the star
distribution taking advantage from the knowledge of the exact star
positions even in the innermost central region. We applied the
procedure described in Montegriffo et al. (1995) which computed
$C_{\rm grav}$ by simply averaging the $\alpha$ and $\delta$
coordinates of stars lying in the PC camera of the HST catalog.
$C_{\rm grav}$ is located at $\alpha_{\rm J2000} = 19^{\rm h}\,
10^{\rm m}\, 52\fs04,~\delta_{J2000} = -59\arcdeg\, 59\arcmin\,
04\farcs64$ with a typical $1\sigma$ uncertainty of $0\farcs 5$
in both $\alpha_{\rm J2000}$ and $\delta_{J2000}$, corresponding
to about $10$ pixels in the WFPC2/HST images. 

AO89 derived the $C_{\rm grav}$
by using the barycenter of the bright ($V<16$) resolved stars from
ground based observations.  We transformed the center position in
their Figure 3,\footnote{Note that the orientation in both the maps
shown by AO89---Figures 1 and 3, respectively---is incorrectly
reported in their captions since E is on the right and N is at the
bottom.}  into our coordinate system.  It well agrees with our
determination, being located only $\sim 2\arcsec$ S.
 
The coordinates of the cluster center previously reported in the
literature refer mostly to the center of luminosity $C_{\rm lum}$,
which is usually determined by the so-called {\it mirror
autocorrelation} (see Djorgovski 1988). $C_{\rm grav}$ turns out to be
$\sim 10\arcsec$ S and $\sim 2\arcsec$ E of the $C_{\rm lum}$
reported in the Djorgovski (1993) compilation. Such a difference is
not surprising---similar offsets have been found in other clusters
(see for example 47 Tucanae---Calzetti et al. 1993, Montegriffo et
al. 1995; M80---Ferraro et al. 1999a).
  
Figure~\ref{map} shows the central $20\arcsec\times 20\arcsec$
of the cluster with respect to the $C_{\rm
grav}$ determined in this work (marked in Figure with a large cross
at $(0,0)$).  The center position by AO89 ({\it small cross} labeled
with AO) and by Djorgovski (1993; {\it small cross} labeled with D)
are also indicated.

Pooley et al. (2002) noticed that the highest concentration of
X-ray sources detected by {\it Chandra} in a 30 ksec pointing toward
NGC~6752 appeared surprisingly displaced with respect to the
optical center by Djorgovski (1988). On the contrary, Figure~\ref{map} shows
that our value of $C_{\rm grav}$ lies close to the barycenter of the
nine more central objects (marked in Figure \ref{map} with a large
cross labeled with XRS at coordinates ($\alpha_{\rm J2000} = 19^{\rm h}\,
10^{\rm m}\, 51\fs82, ~\delta_{J2000} = -59\arcdeg\, 59\arcmin\,
03\farcs84$), reconciling their
projected positions with an almost spherical distribution.

\section{The surface brightness and star density profiles}
\label{profiles} 
By using the combined data set shown in Figure \ref{CMD} we computed
star density and surface brightness profiles applying the standard
procedure fully described in Calzetti et al. 1993 (see also Ferraro et
al. 1999a, Paltrinieri et al. 2001). The entire photometric sample has
been divided in 35 concentric annuli centered on $C_{\rm grav}$, 
spanning a spatial range from $0\arcsec$ to
$27\arcmin$; each annulus has been in turn split in a number of
sub-sectors (generally octants or quadrants depending on the shape of
the FoV covered by the HST and WFI fields). The surface brightness of
each sub-sector has been evaluated as the sum of the luminosity of all
the stars lying in it, divided by the area (expressed in ${\rm
arcsec^2}$).  The average and the rms over the brightness of the
sub-sectors determine the value and the uncertainty of the surface
brightness of an annulus.  However, evaluating the brightness over
small regions can suffer by large fluctuations due to small number
statistics of few bright giants. Hence, we computed three radial
profiles, removing the stars brighter than $V=12,~ V=13,~ V=14$,
respectively.  Figure~\ref{brightness} shows that the overall
structure of the profile does not change with the adopted magnitude
limit.  In contrast, the rms between sub-sectors significantly
decreases once the brightest stars are excluded.  The brightness
profile by AO89 is shown (as open circles) for comparison in the lower
panel of Figure~\ref{brightness}. The agreement with the results of
this work is excellent, although their profile is slightly fainter in
the very central region, probably due to the incompleteness of their
sample. This might be due to the fact that removal of the giants can
lead to an unintentional subtraction of the contribution of close,
fainter stars, which are often unresolved in ground-based
observations.

We can take advantage from the fact that we resolve the stars even in
the innermost region of the cluster and directly produce a star
density profile. As noted by LCG95, a star density profile is not
affected by the statistical fluctuations caused by the bright stars,
and it is the most robust mean for determining the cluster structure
parameters.  Once the photometric sample has been divided in annuli
and sub-sectors, the number of stars lying within each sub-sector has
been counted and averaged. Star density has been eventually obtained
by dividing the average star number by the area of the sub-sector
(expressed in ${\rm arcsec^2}$). 
The star density profile values obtained from this
procedure are listed in Table 1.
 This is the most complete and extended density
profile ever published for this cluster, since it samples the cluster
population from the very inner region out to $r\sim 27\arcmin$ from
the cluster center.

\section{Modeling the radial density profile}
\label{modeling}
Usually GGCs are considered core-collapsed or not depending on how 
well their radial distribution of stars is fitted by King (1966) 
profiles  (Trager et al. 1995). These models are characterized by two 
parameters, the core radius, $r_c$, and the tidal radius, $r_t$, or, 
alternatively, the concentration  $c=\log(r_t/r_c)$.  As stated by 
Meylan \& Heggie (1997, hereafter MH97)
a more general rule is that all clusters with a concentration 
parameter $c>2$--2.5 can be considered as collapsed, on the verge 
of collapsing or just beyond the collapse. 
 
In order to reproduce the observed profile, we obtained the
surface density by projecting the star density from a standard
isotropic, single-mass King-model derived using the code described in
Sigurdsson \& Phinney (1995).\footnote{In our earlier work we have
used an analytic representation of King profiles. Parameter values
derived from the analytic fits differ somewhat from those derived
using the current scheme.}  The result is shown in
Figure~\ref{dens1}. As can be seen, a King model which properly fits
the outer region significantly overestimates the star density over an
extended inner region ($5$--$20\arcsec$). This fit closely resembles
the result obtained by LCG95 who fitted the $U$-band surface
brightness profile for $r<120\arcsec$ with a $c=2$ King model (see
their Figure {\it 2r}).  On the basis of that fit, they concluded that
the cluster is not required to be in a post-core collapse state
because the model did not differ from the data in a statistically
significant way.  However, there were aspects of the fit which led us
to explore other possibilities. These included the bad fit in the
region $5 < r < 20\arcsec$ and the fact that even in the outer parts
there are regions where the data line systematically above or below
the observed points. We suspected that unmodeled effects were
degrading this fit.

For this reason, we searched for alternative solutions.  As
shown in Figure \ref{density} the density profile can be well fit by a
dual King model. The outer cluster ($r>10\arcsec$) is well fit by a
model with $r_c=28\arcsec$ and $c=1.9$. The observed counts are
significantly in excess of this model within the central
$8\arcsec$. In this innermost region a King model with $c=2.1$ and
$r_c=5.7\arcsec$ fits the data, but lies significantly below all of
the points with $r>10\arcsec$. A dual King model does not represent a
detailed equilibrium of cluster structure. However, it is consistent
with the scenario in which the central regions have evolved away from
a global King model that now only characterizes the outer regions.

This anomalous structure of the innermost radial profile of NGC~6752
has been evident since the very first studies (Da Costa 1979). The
$U$-band data presented by DK86 showed a sharp shoulder in the radial
profile and prompted the authors to classify NGC~6752 as a
core-collapsed cluster. A few years later, AO89 modeled their
detailed $V$-band profile with a power law in the region between
$3\arcsec$ and $60\arcsec$ from the center, noting a significant
flattening for $r<3\arcsec$.  


LCG95   used also a  
modified power-law (see their eq. 3, also reported in
Figure \ref{powerlaw}), in order to   test the presence of
resolved cores.   We used
the inner part of the star density profile of Figure
\ref{density} to repeat an analysis similar to that
presented by LCG95. Besides a normalization factor, their
expression is a function of two free parameters: the
power-law index $\alpha$ and the scale-length of the core
$r_0$, which can be easily converted in the usual $r_c$
(see their eq. 2).  As can be seen in Figure \ref{powerlaw}
a proper fit (with $\chi^2/{\rm d.o.f}=0.28$ for $8$ d.o.f)
can be obtained, yielding $r_0=3\farcs1\pm 1\farcs4$
(corresponding to a core radius $r_c=5\farcs2\pm 2\farcs
4$) and $\alpha=-1.05\pm 0.05$ ($1\sigma$ errors). These
values are consistent with the previous results by  
LCG95 ($r_c=6\farcs 7 \pm 1\farcs 9$  and $\alpha=-0.97\pm
0.15$) and by AO89, who estimated a radius of the central
plateau of $\sim 3\arcsec$ (from the analysis of their
images) and $\alpha=-0.95$ (from the brightness density
profile). At a distance of $D=4.3\pm 0.4$ kpc (Ferraro et
al. 1999b), the core radius we inferred  corresponds to a
physical dimension of $0.11\pm 0.05$ pc.

\section{The dynamical status of NGC~6752}
\label{dyn}
 
The significant deviation of the star number density profile from a
canonical King model is a clear indication that the innermost region
NGC~6752 has experienced (or is experiencing) a collapse
phase. Unfortunately, no similarly unambiguous signature is available
for differentiating the in- and post-core collapse state (Meylan \& Heggie
1997).  However, in favorable cases the phase of the collapse can be
evaluated from indirect evidence.

Our results solidify the earlier
suggestions of post-core collapse bounce made by AO89.  Post-core
collapse clusters are expected to undergo large amplitude oscillations
in core size due to the gravothermal instability of collisional
systems (e.g. Cohn et al. 1991). The oscillating core spends most of
the time at near maximum size and a radial extension of $0.11$ pc is
consistent with the maximum radius of the core predicted by the models
of post-core collapse bounce. The parameter most commonly used in
theoretical studies is the ratio of core radius to half-mass radius,
$r_c/r_h$.  Using our value of $r_c = 5.2\arcsec$ and the half-mass
radius of $115\arcsec$ from Trager et al (1995), we find a ratio
$r_c/r_h = 0.045$.  For comparison, the multi-mass Fokker-Planck
models for the post-collapse evolution of M15 presented by Grabhorn et
al (1992, see their Fig. 5) reach a comparable value of $r_c/r_h =
0.034$ during the maximally expanded state of the most extreme core
bounces.  The inclusion of primordial binaries in Monte Carlo
simulations by Fregeau et al (2003) results in an even wider range of
predicted $r_c/r_h$ in the post-collapse phase.

Furthermore, we have found that there is an
intermediate region in the composite density profile of NGC~6752 which
is well represented by a power law profile with a slope $\alpha \sim
-1,$ compatible with the steepness predicted by single mass models of
expanding bouncing cores (LCG95).

It is also worth noting that NGC~6752 has
apparently retained a substantial primordial binary population
(Rubenstein \& Bailyn, 1997); these binaries may play an important
role in supporting the core and delaying the core-collapse event.   
In this respect, Ferraro et al (1999a) have suggested that some species
originated from binary evolution could be used as possible tracers of
the cluster dynamical evolution.  In particular, the large BSS
population recently found in M80 by Ferraro et al (1999a) might be the
signature of a transient dynamical state, during which stellar
interactions are delaying the core-collapse process leading to an
exceptionally large population of {\it collisional}-BSS. On the other hand the
BSS population found in the central region of NGC~6752 is small
(Ferraro et al 2003, in preparation) perhaps indicating that NGC~6752 is
in a different dynamical evolutionary state than M80. Maybe the
binary population in NGC~6752 has not been ``burned out''  producing
collisional BSS while that in M80 has.

\section{The interpretation of the MSPs accelerations}
\label{msp}

NGC~6752 hosts 5 known millisecond pulsars (D'Amico et al. 2001a;
D'Amico et al. 2002). The positions in the plane of the sky of three
of them (PSRs B, D and E, all isolated pulsars) are close to the
cluster center, as expected on the basis of mass segregation in the
cluster. In particular PSR-B and E show large {\it negative} values of
$\dot{P}$, implying that the pulsars are experiencing an acceleration
with a line-of-sight component $a_l$ directed toward the observer
and a magnitude 
significantly larger than the positive component of $\dot{P}$ due
to the intrinsic pulsar spin-down (see e.g. Phinney 1992). What is
the origin of such acceleration?  Given the location of NGC~6752 in
the galactic halo and knowing its proper motion (Dinescu et al. 1999),
it is possible to calculate the contributions to $\dot{P}$ due to
centrifugal acceleration, differential galactic rotation and vertical
acceleration in the Galactic potential, all of them resulting
negligible (D'Amico et al. 2002).  Hence, the remaining plausible
explanations of the observed negative $\dot{P}$ are: the
accelerating effect of the cluster gravitational potential well or
the presence of some close perturbator(s) exerting a
gravitational pull onto the pulsars. Taking advantage from 
the new results presented in this paper we discuss in the follow
viability and implications of these two possibilities.

\subsection{Case (i): Overall effect of the GC potential well}
\label{case1}

The hypothesis that the line-of-sight acceleration of the
MSPs with negative $\dot{P}$ is dominated by the cluster
gravitational potential has been routinely applied to many
globulars. In particular from this assumption a
lower limit to the mean projected mass-to-light ratio in
the $V$-band ${\cal{M}}/{\cal{L}}_V$  in the central region
of M15 (Phinney 1992) and 47 Tucanae (Freire et al. 2003),
yielded ${\cal{M}}/{\cal{L}}_V$$\gapp 3$ and $\gapp 0.7$
respectively.  Following Phinney (1992), a lower limit to
$\cal{M}/{\cal L}_V$ in the inner regions of NGC~6752 is
given by

\begin{eqnarray*}
 & \left|\frac{\dot{P}}{P}(\theta_{\perp})\right| <
\left|\frac{a_{l,max}(\theta_{\perp})}{c}\right| \simeq & \\
 & \simeq 1.1\frac{G}{c}\frac{M_{cyl}(<\theta_{\perp})}{\pi
D^2\theta^2_{\perp}}=5.1\times 10^{-18} \frac{\cal{M}}{{\cal
L}_V}\left(\frac{\Sigma_{V}(<\theta_{\perp})}{10^4~{\rm L_{V\odot}
pc^{-2}}}\right){\rm s^{-1}},
\label{eq:Sigma}
\end{eqnarray*}
\noindent where $\Sigma_{V}(<\theta_{\perp})$ is the mean surface
brightness within a line of sight subtended by an angle
$\theta_{\perp}$ with respect to the cluster center and
$M_{cyl}(<\theta_{\perp})$ is the mass enclosed in the cylindrical
volume of radius $R_{\perp}=D\theta_{\perp}.$ This equation
holds to within $\sim$10\% in all plausible cluster models and
is independent of cluster distance, except for the effects of
extinction.  Since $E(B-V)$ is very small for NGC~6752 (= 0.04
according to Harris 1996) the latter is a negligibe affect for
this cluster.

Using the observed
$\dot{P}/P$ of PSR-B and PSR-E (D'Amico et al. 2002), combined
with our accurate determinations of $C_{\rm grav}$ (the cluster center
of gravity) and $\Sigma_{V}(<r)$ (the mean surface brightness radial
profile), it turns out that ${\cal{M}}/{\cal{L}}_V \gapp 6$--7 (see Figure
\ref{MtoL}) for the case of NGC~6752. D'Amico et al. (2002) obtained a
slightly larger ${\cal{M}}/{\cal{L}}_V \gapp 10$ using published
values of $C_{\rm grav}$ and $\Sigma_{V}(<r)$ derived from medium
resolution ground based observations only. The difference between the
two estimates is mainly due to our new position of the cluster
center of gravity. Despite a residual uncertainty $\sim 0\farcs 7$ on
$C_{\rm grav},$ under the hypothesis that the line-of-sight
acceleration of PSR-B and PSR-E are entirely due to the cluster
gravitational potential, a lower limit of ${\cal{M}}/{\cal L}_V\gapp
5.5$ can be firmly established. It is obtained assuming that the two
millisecond pulsars were just symmetrically located (and hence at the
minimum projected distance) with respect to the actual center of
gravity.

The sample of the core collapsed clusters shows typical values of the
projected central mass-to-light ratio in the interval 2--3.5 (Pryor \&
Meylan 1993), although larger ${\cal{M}}/{\cal L}$ ratios can
be obtained when a Fokker--Planck model fit is used (e.g. the case of
M15 Dull et al. 1997, 2003).  If we take ${\cal{M}}/{\cal{L}}_V \gapp
3$ the expected total mass located within the inner
$r_{\perp,B}=0.08$ pc of NGC~6752 (equivalent to the projected
displacement of PSR-B from $C_{\rm grav}$) would be $\sim
1200$--$2000\,{M_\odot}.$ On the other hand, the observed
${\cal{M}}/{\cal{L}}_V$$\sim 6$--7 implies the existence of further
$\sim 1500$--$2000\,{M_\odot}$ of low-luminosity matter segregated
inside the projected radius $r_{\perp,B}.$ This extra amount of mass
could be constituted by a relatively massive black hole, like the
$\sim 1700^{+2700}_{-1700}\,{M_\odot}$ black hole in the center of the
globular cluster M15 recently proposed by Gerssen et
al. (2003). However, the results presented in \S 4 and \S 5 show that
there are significant differences in the kinematics and mass
distributions of the central regions of M15 and NGC~6752 (we note here
that because the distance to NGC~6752 is less than half that to M15,
its inner core is more easily studied): {\it (1.)} HST imaging of M15
(Guhathakurta et al. 1996---WFPC2; Sosin \& King 1997---FOC) shows no
evidence for a compact core, at variance with our observations of the
core of NGC~6752; {\it (2.)} The derived stellar density profiles of
M15 have power law slopes consistent with $\alpha = -0.75$ (expected
from single mass models with a dominant central black hole; LCG95).
If a $\gapp 10^3\,{M_\odot}$ black hole resides in the inner region of
NGC~6752, its gravitational influence would extend more than $\sim
2\arcsec$ from the center of the cluster and probably produce a
central power-law cusp, which we do not observe.

A very high ${\cal{M}}/{\cal{L}}_V \sim 6$--7 could be also due to
central concentration of dark remnants of stellar evolution, like
neutron stars (NSs) and heavy $\sim 1.0\,{M_\odot}$ white dwarfs (WDs)
(as also proposed for M15 by Dull et al. 1997, 2003 and by Baumgardt
et al. 2003). In this case, one can constrain the initial mass
function (IMF) and/or the neutron star retention fraction $f_{\rm
ret}$ in NGC~6752. Based on the
current population of turn-off stars (in the mass interval
$0.6$--$0.8\,{M_\odot}$), the estimated number of upper main sequence
initially present in NGC~6752 is $\sim 4000$ (D'Amico et
al. 2002). This assumes a Salpeter-like IMF ($\alpha_{\rm IMF}=2.35$
which is consistent with that measured by Ferraro et al. 1997b).  If
the low-luminosity matter observed in the central 0.08 pc were
entirely due to $\approx 1300$ NSs of $1.4\,{M_\odot}$, then $f_{\rm
ret}\sim 30$\% (a reasonable value for collapsed
clusters [Drukier
1996]). Alternatively, ${\cal{M}}/{\cal{L}}_V \sim 6$--$7$ can
be explained by a Salpeter IMF if $\gapp 20$\% of the total population
of heavy $1.0\,{M_\odot}$-WDs sank into the NGC~6752 core during the
cluster dynamical evolution. Either scenario must also be consistent
with the observed shape of the star density profile.

According to Cohn (1985), during the core collapse phase
the surface density slope for the most massive component is expected
to be $\alpha=-1.23$ rather than the projected isothermal slope of
$\alpha=-1.0$.  In the central cusp that forms during core collapse,
the surface density slope of a component of stellar mass $m$ is given
approximately by:

$$\alpha = -1.89 {{m}\over{m_d}} +0.65$$

\noindent where $m_d$ is the stellar mass of the dominant component.
If the luminosity profile is dominated by turnoff stars of mass $m =
0.8 M_{\odot}$ and has a slope of $\alpha = -1.05$, then the implied
mass of the dominant, nonluminous component should be $m_d =
0.89M_{\odot}$ , i.e. somewhat more massive than the adopted turnoff
mass.  This argument does suggest that the central gravitational
potential is not likely to be dominated by a large number of neutron
stars, but heavy white dwarfs still remain a possibility.

Velocity dispersions provide a further constraint on the nature of the
cluster potential well. The stars dominating the dynamics in the inner
part of the cluster should have (see eq. 3.5 of Phinney 1992) a
1-dimensional central velocity dispersion $\sigma_{v,0}\gapp
9$--$10\,{\rm km\, s}^{-1}$. This is compatible both with the very
wide published $2\sigma$ interval for the $\sigma_{ v,0}$ of NGC~6752
($2.1$--$9.7\,{\rm km\, s}^{-1}$, Dubath, Meylan \& Mayor 1997) and
with preliminary proper motion measurements of stars in the central
part of NGC~6752, which would suggest a significantly higher
one-dimension velocity dispersion $\sigma_{ v,0}\sim 12 \,{\rm km\,
s}^{-1}$ and the existence of strong velocity anisotropies (Drukier
et al. 2003){\footnote{If confirmed, such anisotropies would
support the hypothesis of a relatively massive, probably binary black
hole}}. Recent Fabry-Perot spectroscopy of single stars in NGC~6752
shows a flat velocity dispersion profile with typical one-dimension
velocity dispersion $\sigma\sim 6$--7$\,{\rm km\, s^{-1}}$ within the central
$1\arcmin$ (Xie et al. 2002). This is also marginally compatibile (given the
$\sim 10$\% uncertainty of the formula 3.5 of Phinney 1992).

\subsection{Case (ii): Local perturbator(s)}
\label{case2}

We here explore the alternate possibility
that the acceleration imparted to PSR-B and PSR-E are due
to some local perturbator. 

NGC~6752 is a highly concentrated cluster and its core could host
$n_*\gapp 10^6$ stars per cubic pc (assuming an average stellar mass
of $\overline{m}\sim 0.5\,{M_\odot}$). Hence close star-star
encounters are a viable possibility. In order to produce the
line-of-sight acceleration $|a_{l}|=c|\dot{P}/P|=2.9\times
10^{-6}\,{\rm cm\, s}^{-2}$ seen in PSR-B (or PSR-E) a passing-by
star of mass $\overline{m}$ must approach the pulsar at $s\le
(G\overline{m}|a_{l}|^{-1})^{1/2}=0.0015~(\overline{m}/0.5\,{
M_\odot})^{1/2}$ pc. An upper limit to the probability of occurrence
of a suitable close encounter can be roughly estimated as $\sim
s^3n_*=3.7\times 10^{-3}(\overline{m}/0.5\,{M_\odot})^{3/2}.$ Although
this figure is not negliglible, the need of having two different
canonical stars independently exerting their gravitational pull onto
PSR-B and PSR-E, makes the joint probability of such configuration
suspiciously low, $\lapp 10^{-5}$. Rather than being randomly placed
could the perturber be a binary companion to the MSPs?  For such a
companion not to have been aleady discovered by pulsar timing analysis
would require an orbital period $P_b\gapp 20$ yr. Survival of such a
wide binary is quite problematic in the core of a dense cluster like
NGC~6752---indeed no binary pulsar with $P_b>3.8$ day has been detected
in collapsed clusters to date. Seeing two such systems in NGC~6752
appears extremely unlikely.

\begin{figure}
\plotone{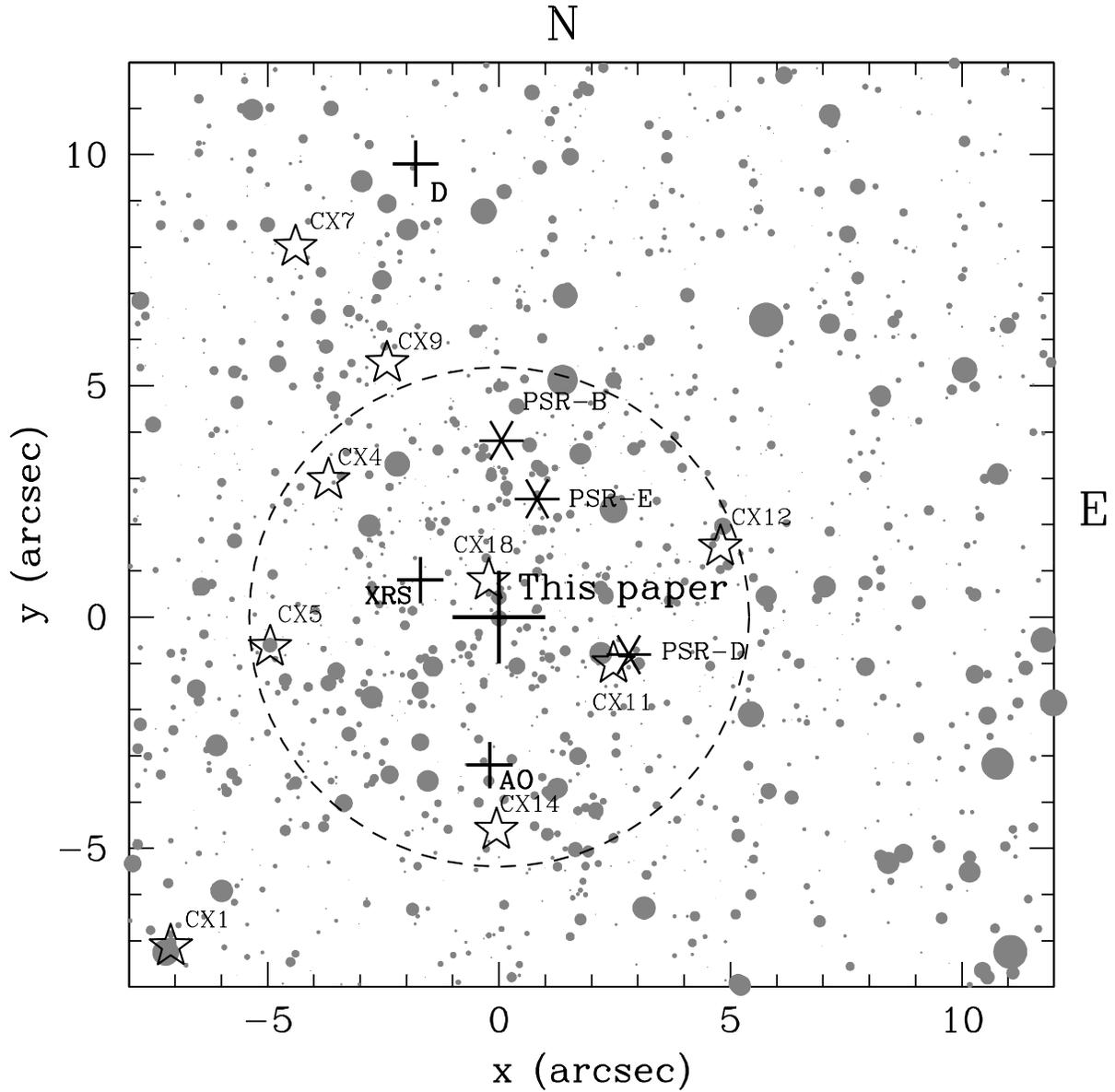} 
\caption{\label{map}
\footnotesize{Computer map of the inner 
$20\arcsec\times20\arcsec$ region of the cluster. The {\it large cross} at
coordinate $(0,0)$ indicates the adopted $C_{\rm grav}$.  The {\it
small crosses} are previous determinations of the cluster center: the
$C_{\rm lum}$ by Djorgovski (1993) is labeled with D and the $C_{\rm
grav}$ by AO89 (obtained considering only bright stars at $V<16$) is
labeled with AO.  The nominal position of the 9 more central {\it
Chandra} X-ray sources (Pooley et al. 2002) and of 3 MSPs are shown as
{\it empty star symbols} and {\it asterisks}, respectively.  The
barycenter of the 9 X-ray sources is also shown as a {\it small cross}
and labeled with XRS. The {\it dashed} circonference encircles the core
of the cluster.}}
\end{figure}

\begin{figure}
\plotone{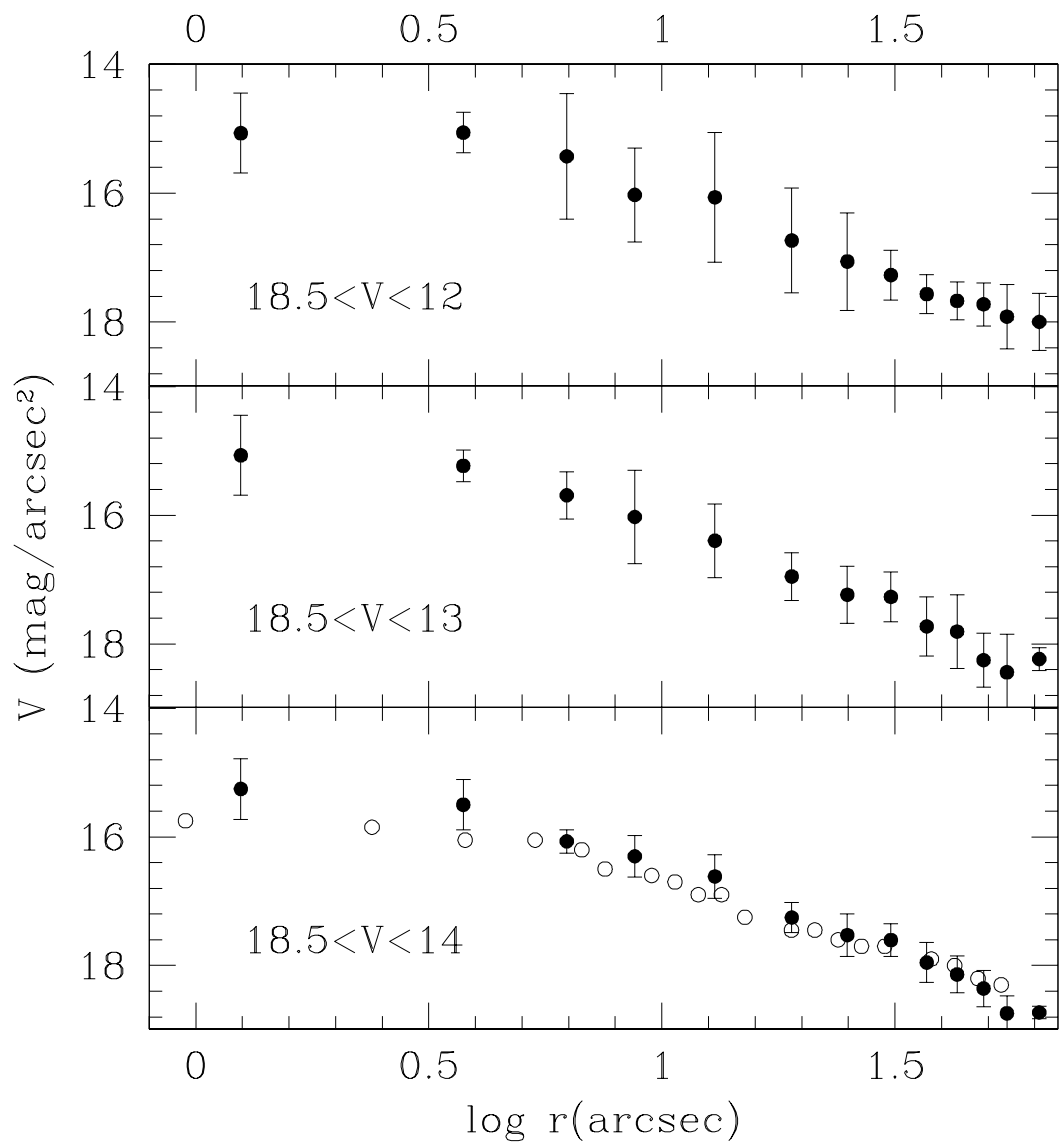} 
\caption{\label{brightness}
\footnotesize{Radial brightness profiles computed removing the stars
brighter than $V=12,~V=13,~V=14$, respectively, the faint limit always
being $V=18.5$ (see Figure \ref{CMD}).  From the data of the lower
plot, the best estimate of the central surface brightness is
$15.07~{\rm mag\,arcsec^{-2}}$.  Correction for extinction (using
$E(B-V)=0.04$ , Harris 1996) leads to to $3.67\times 10^4~{\rm
L_{V\odot}\, pc^{-2}}.$ In the lower panel the data from AO89 ({\it
open circles}) are shown for comparison.}}
\end{figure}

\begin{figure}
\plotone{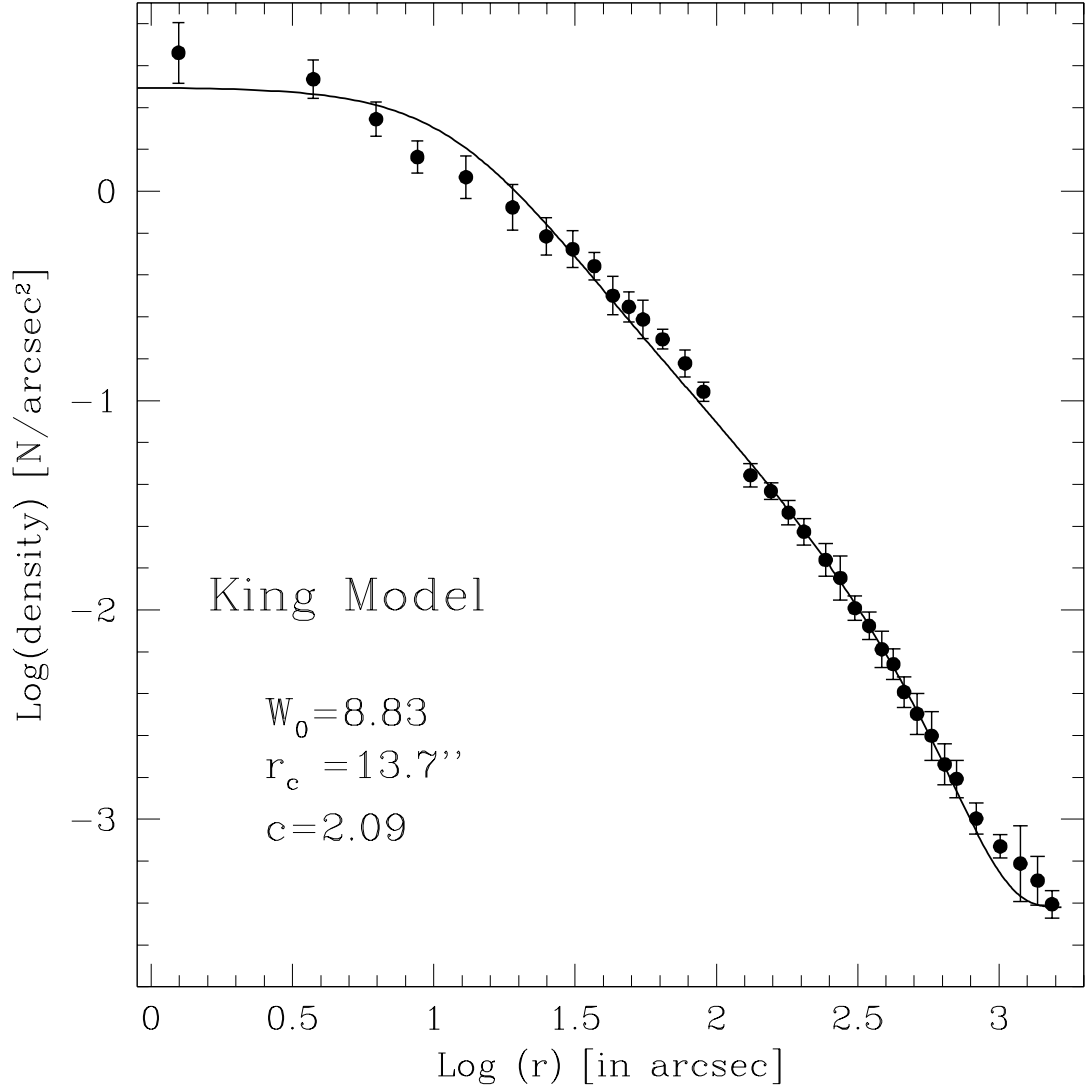} 
\caption{\label{dens1}
\footnotesize{Observed radial density profile with respect to the the
adopted $C_{\rm grav}$.  The solid line is the best fit King model
($r_c=13\farcs 7$ and $c=2.1$) to the observed density
profile over the entire extension. Note that the adopted profile
significantly 
understimates the star counts in the region between
$3\arcsec$ and $20\arcsec$. The adopted value for the
parameter $W_0$ (the central potential parameter defined by
King 1996) is also reported.}}
\end{figure}

\begin{figure}
\plotone{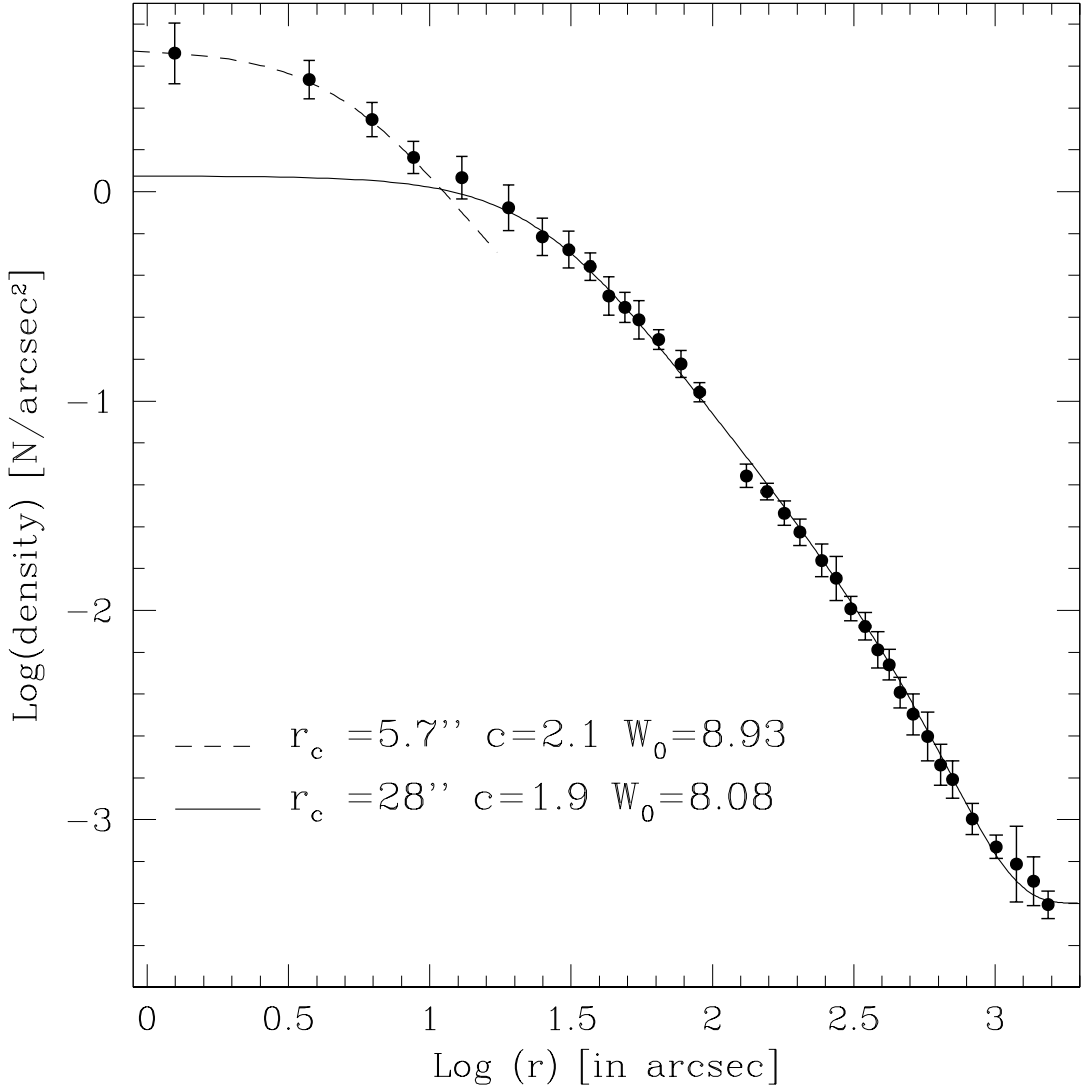} 
\caption{ \label{density}
\footnotesize{Observed radial density profile with respect to the the
adopted $C_{\rm grav}$.  The solid line is the best fit King model
($r_c=28\arcsec$ and $c=1.9$) to the outer points (that is, for
distance larger than $10\arcsec$ from the cluster center).  The King
model has been combined with a constant star density of $ 0.25~{\rm
stars\,arcmin^{-2}}$ in order to account for the flattening of the
density profile in the extreme outer region ($r>16\farcm5$) of the
cluster. The dashed line is the best fit King model
to the 4 innermost points.}}
\end{figure}

\begin{figure}
\plotone{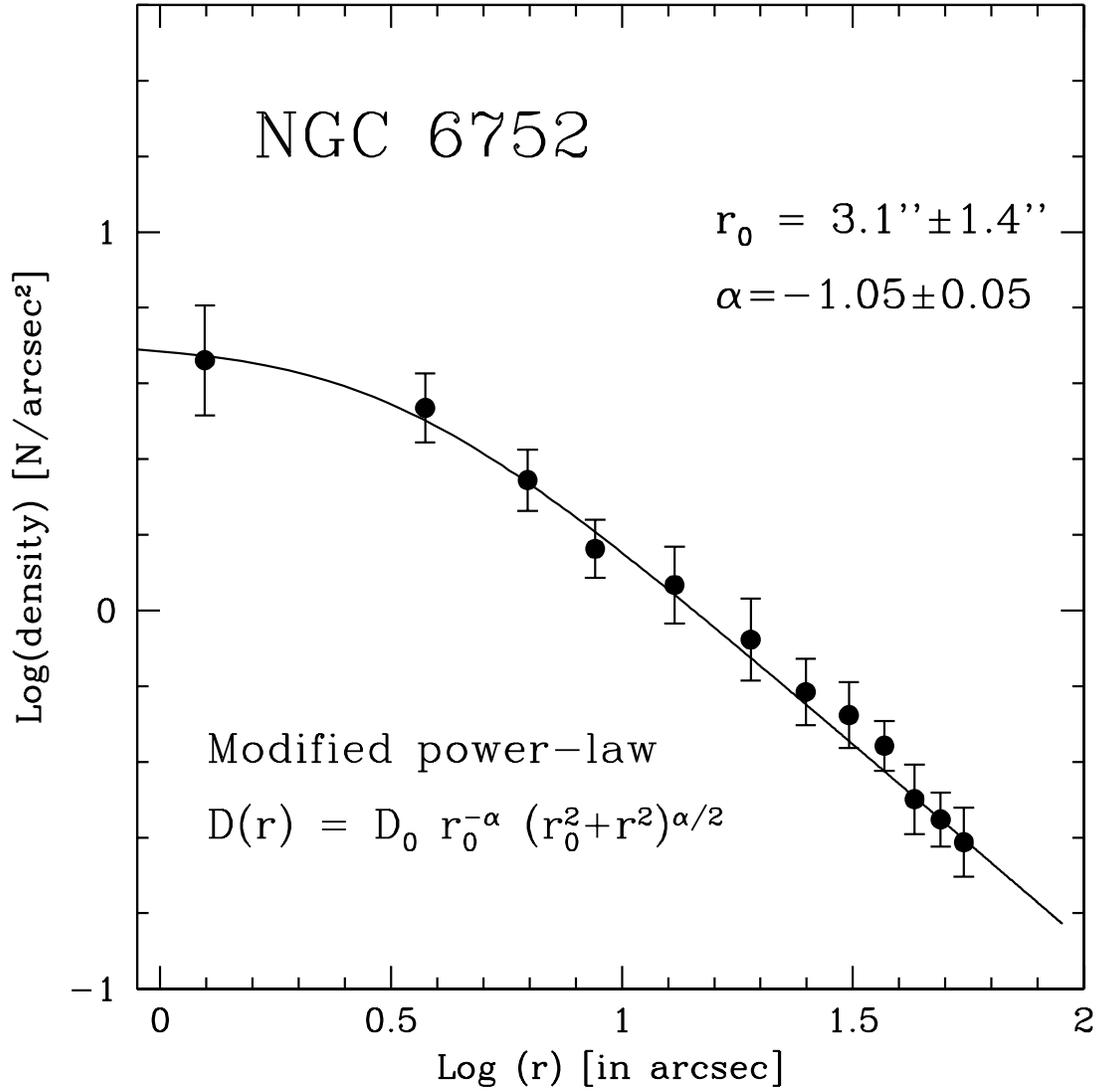} 
\caption{\label{powerlaw} 
\footnotesize{The central part ($r<1\arcmin$) of the radial density
profile of NGC~6752.  A modified power-law (from LCG95) has been used
to fit the data. The solid line represents the best fit solution, whose
parameters are also marked. The reduced $\chi$-squared is $0.28$.
}}
\end{figure}

\begin{figure}
\plotone{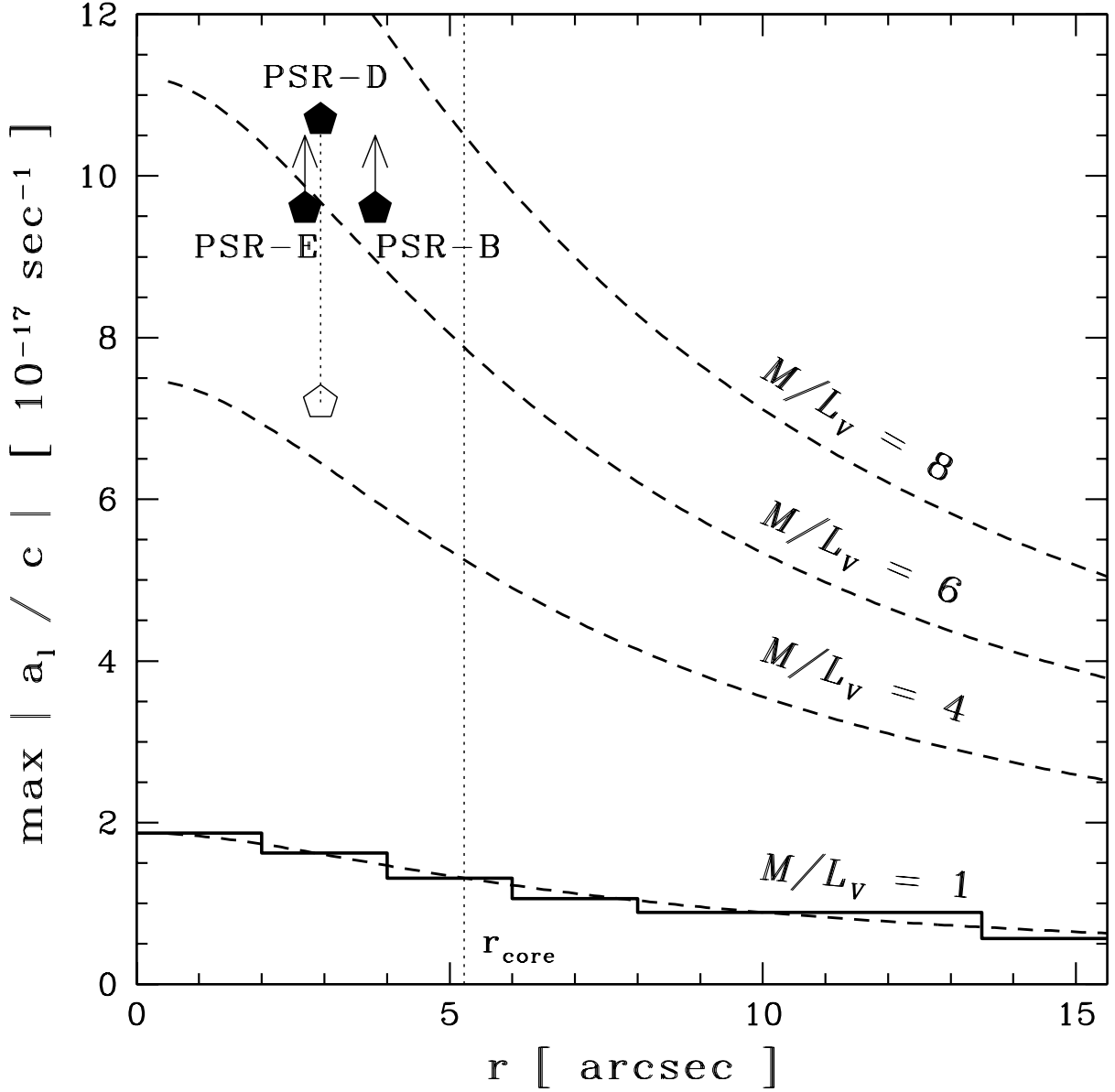} 
\caption{\label{MtoL} 
\footnotesize{Maximum line-of-sight acceleration $|a_{{l}_{max}}/c| =
|\dot{P}/P|$ versus radial offset with respect to the center of
NGC~6752. The histogram represents the prediction based on the star
density profile of \S\ref{profiles} (normalized to the central surface
brightness in the $V$-band obtained from the data of
Figure~\ref{brightness}) assuming a unity mass-to-light ratio.  The
dashed lines are analytical fits to the optical observations, labeled
according to the adopted mass-to-light ratio.  The measured values of
$\dot{P}/P$ (filled pentagons, D'Amico et al. 2002) in the two MSPs
with negative $\dot{P}$ (PSR-B and E) can be reproduced only for
${\cal{M}}/{\cal L}_V\gapp 6-7.$ The open pentagons show our best
guessed range of maximum $|a_l/c|$ for PSR-D: the upper value is
calculated assuming a negligible intrisic positive $\dot P_{sd}$; the
lower value is estimated taking into account intrinsic $\dot P_{sd}$
from the observed scalings between X-ray luminosity and spindown power
for MSPs (see D'Amico et al. 2002 and reference therein).  Given the
relative large uncertainty, the value of $|a_l/c|$ it is not further
taken into account in the discussion.}}
\end{figure}

One may wonder if a {\it single} object, significantly more massive
than a typical star in the cluster, could simultaneously produce the
accelerations detected both in PSR-B and PSR-E.  Recently, Colpi,
Possenti \& Gualandris (2002) suggested the presence of a binary
black-hole (BH) of moderate mass ($M_{\rm bh+bh}\sim 100$--$200\,{
M_\odot}$) in the center of NGC~6752 in order to explain the
unprecedented position of PSR-A, a binary millisecond pulsar which
is far away from the cluster center (D'Amico et al 2002).  As shown in
Figure \ref{map}, the projected separation of PSR-B and PSR-E is only
$d_{\perp}=0.03$ pc.  A binary BH of total mass $M_{\rm bh+bh}$,
approximately located in front of them within a distance of the same
order of $d_{\perp}$, could be accelerating both the pulsars without
leaving any observable signature on the photometric profile of the
cluster.  As the BH binarity ensures a large cross section to
interaction with other stars, the recoil velocity $v_{\rm rec}$ due a
recent dynamical encounter could explain the offset position (with
respect to $C_{\rm grav}$) of the BH. However this scenario suffers of
a probability at least as low as that of the previous two: placing the
black hole at random within the core gives roughly a 1\% chance that
it would land in a location where it 
would produce the observed pulsar
accelerations and the required 
$v_{\rm rec}\gapp 4$--$5\,{\rm km\, s}^{-1}$ is
also at the upper end of the expected distribution of $v_{\rm rec}$
for an intermediate mass black-hole in NGC~6752 (Colpi, Mapelli \&
Possenti 2003, in preparation).  

\vskip 1.0truecm In summary, on the basis of the available data, the
accelerations shown by PSR-B and PSR-E can be easilier accounted by
the usually adopted hypothesis {\it (i)} (which could explain also the
large positive $\dot{P}$ of PSR-D if it were not instrinsic), although
the nature of the required extra amount ($1500$--$2000\,{M_\odot}$) of
low luminosity mass still remains puzzling.  The existence of local
perturber(s) of the pulsar dynamics (case {\it (ii)}) is a distinct
possibility. While it seems extremely unlikely purely on the basis of
the high value of $\dot{P}/P$ measured in PSR-B and PSR-E, there are
other indications of a binary low-mass BH. In particular, it would
explain: {\it (a)}~the absence of any cusp in the radial density
profile, {\it (b)}~the flat velocity dispersion profile (Xie et
al. 2002), {\it (c)}~ and {\it (d)}~the ejection of PSR-A and PSR-C in
the cluster outskirts (Colpi, Possenti \& Gualandris 2002). However it
is admittedly {\it ad hoc} hypothesis, requiring a fine tuned
scenario.

Clearly additional information must be collected.  A longer baseline
for timing measurements will allow to better constrain the presence of
companions in very large orbits around PSR-B and PSR-E. In
particular it will permit to derive (at least) upper limits on the
second derivative of the pulsars' spin period, which is more
influenced by by-passing stars rather than by the cluster potential
well (Phinney 1992). Similarly, the (single massive or binary
intermediate mass) black-hole hypotheses could be better investigated
with spectroscopic determination of the dispersion velocity of the
stars located in the pulsars' neighborhood, likewise with star density
counts reaching higher magnitudes (hence exploiting a larger sample of
objects) and better spatial resolution (thus probing the innermost
$1\arcsec$ of the cluster and the pulsars' surroundings).

\acknowledgements 
We thank Phyllis Lugger for useful comments and suggestions
that significantly improved the presentation of the paper.
We also thank P. Montegriffo for assistance with the
astrometry procedure, E. Pancino for the pre-reduction
of the WFI images and S. Sigurdsson for allow us to use his
code to compute King model surface density profile.
Bingrong Xie kindly provided results prior to publication.
Financial support to this research is provided by the
Agenzia Spaziale Italiana (ASI) and the {\it Ministero
dell'Istruzione, dell' Universit\`a e della Ricerca}
(MIUR).  The GSCII catalog was produced by the Space
Telescope Science Institute and the Osservatorio
Astronomico di Torino. RTR is partially supported by STScI
grant GO-8709 and NASA LTSA grant NAG5-6403.

\newpage
\newdimen\minuswidth    
\setbox0=\hbox{$-$}
\begin{deluxetable}{lcccllccc}
\scriptsize \tablewidth{14.7cm}  
\tablecaption{\label{t:mjd}
Surface brightness}
\tablecomments{$r1$ and $r2$ are the inner and outer radii
  of each annulus in arcsec.}
\startdata \\
\hline
\hline
       &	&                            &     & &      &      &
                            &     \\
  $r1$  &    $r2$  &  $Log(\frac{N}{{\rm arcsec^2}})$ & err & & $r1$  &  $r2$  & 
  $Log(\frac{N}{{\rm arcsec^2}})$ & err \\
       &	&                            &     & &      &      &
                            &     \\
\hline        
  0.0 &   2.5 &  0.66121 & 0.14539 &  &  192.0 &  216.0 & -1.62649 & 0.06390 \\
  2.5 &   5.0 &  0.53520 & 0.09093 &  &  228.0 &  259.0 & -1.76117 & 0.07847 \\
  5.0 &   7.5 &  0.34446 & 0.08111 &  &  259.0 &  290.0 & -1.84727 & 0.10556 \\
  7.5 &  10.0 &  0.16290 & 0.07666 &  &  290.0 &  328.0 & -1.99170 & 0.05829 \\
 10.0 &  16.0 &  0.06712 & 0.10091 &  &  328.0 &  366.0 & -2.07631 & 0.06523 \\
 16.0 &  22.0 & -0.07693 & 0.10896 &  &  366.0 &  404.0 & -2.18818 & 0.08721 \\
 22.0 &  28.0 & -0.21536 & 0.08836 &  &  404.0 &  442.0 & -2.25923 & 0.07283 \\
 28.0 &  34.0 & -0.27670 & 0.08735 &  &  442.0 &  480.0 & -2.39233 & 0.07300 \\
 34.0 &  40.0 & -0.35778 & 0.06545 &  &  480.0 &  545.0 & -2.49601 & 0.09764 \\
 40.0 &  46.0 & -0.49884 & 0.09170 &  &  545.0 &  610.0 & -2.60130 & 0.11642 \\
 46.0 &  52.0 & -0.55220 & 0.07127 &  &  610.0 &  675.0 & -2.73773 & 0.09854 \\
 52.0 &  58.0 & -0.61254 & 0.09105 &  &  675.0 &  740.0 & -2.80696 & 0.08906 \\
 58.0 &  71.0 & -0.70716 & 0.04729 &  &  740.0 &  920.0 & -2.99633 & 0.07358 \\
 71.0 &  84.0 & -0.82188 & 0.06490 &  &  920.0 & 1100.0 & -3.12938 & 0.05624 \\
 84.0 &  96.0 & -0.9577  & 0.04571 &  & 1100.0 & 1280.0 & -3.21167 & 0.18171 \\
120.0 & 144.0 & -1.35696 & 0.05556 &  & 1280.0 & 1460.0 & -3.29291 & 0.11643 \\
144.0 & 168.0 & -1.43250 & 0.04013 &  & 1460.0 & 1620.0 & -3.40592 & 0.06529 \\
168.0 & 192.0 & -1.53549 & 0.05733 &  &        &	&	   &	     \\

\hline
\enddata
\end{deluxetable}

\newpage

\newpage

\newpage

\newpage

\newpage

\newpage

\newpage

\end{document}